\documentstyle[b98proc,epsfig]{article}
\include{psfig}
\def\Journal#1#2#3#4{{#1} {\bf #2}, #3 (#4)}

\def\NPB{{\em Nucl. Phys.} B}

\def\PLB{{\em Phys. Lett.} B}

\def\PRD{{\em Phys. Rev.} D}

\begin{document}
\title{
COMPTON SCATTERING ON PION AND PION POLARIZABILITIES\footnote{
This work was supported by the
Deutsche Forschungsgemeinschaft (SFB 201)}}
 
\author{L.V. FIL'KOV, V.L. KASHEVAROV}
\address{Lebedev Physical Institute, Moscow 117924, Russia}

\maketitle

\abstracts{
The Compton scattering on a charged pion  
is studied using the dispersion relations. 
Unknown parameters of the 
$\sigma$ meson are found from a fit to the experimental data for 
$\gamma\gamma\to\pi^0\pi^0$ process.} 

In the the present work we investigate the elastic $\gamma\pi^{\pm}$ scattering 
in the energy region up to $\sqrt{s}\simeq 1$GeV (where $s$ is the square 
of the total energy in $\gamma\pi$ c.m.s.).
With this aim we construct dispersion relations (DRs)
at fixed square of the momentum transfer $t$ with one subtraction
for the helicity amplitudes $M_{++}$ and $M_{+-}$ of the elastic $\gamma\pi$ 
scattering. 
\begin{eqnarray}
\lefteqn{Re M_{++}(s,t)=Re \overline{M}_{++}(s=\mu^2,t)+B_{++}+}  \\
 &&\frac{(s-\mu^2)}{\pi}P\int\limits_{4\mu^2}^{\infty}ds'~Im M_{++}(s',t)
\left[
\frac{1}{(s'-s)(s'-\mu^2)}-\frac{1}{(s'-u)(s'-\mu^2+t)}\right] \nonumber
\end{eqnarray}
where $B_{++}$ is the Born term. 
The subtraction function $Re \overline{M}_{++}(s=\mu^2,t)$
is determined with help of the DR at fixed $s=\mu^2$ with one 
subtraction where the subtraction constant is expressed through 
the difference of the pion polarizabilities $(\alpha-\beta)$:
\begin{eqnarray}
\lefteqn{Re \overline{M}_{++}(s=\mu^2,t)=Re M_{++}(s=\mu^2,t)-B_{++}(s=\mu^2,t)=
2\pi\mu (\alpha-\beta) +} \nonumber \\
& &\frac{t}{\pi}P\left\{\int\limits_{4\mu^2}^{\infty}
\frac{Im M_{++}(t',s=\mu^2)~dt'}{t'(t'-t)}-
 \int\limits_{4\mu^2}^{\infty}
\frac{Im M_{++}(s',u=\mu^2)~ds'}{(s'-\mu^2)(s'-\mu^2+t)}\right\} 
\end{eqnarray}

The DRs for the amplitude $M_{+-}(s,t)$ have the same expressions (1) and (2)
with substitutions: $Im M_{++} \to Im M_{+-}$, $B_{++} \to B_{+-}=B_{++}/\mu^2$ and
$2\pi\mu (\alpha-\beta) \to 2\pi/\mu (\alpha+\beta)$.
The DRs for the $\gamma\pi^{\pm}$ scattering are saturated by 
the contributions of the $\rho(770)$, $b_1(1235)$ and $a_2(1320)$ mesons in the $s$ and
$u$ channels and $\sigma$, $f_0(980)$ and $f_2(1270)$ mesons in the $t$ channel.
The parameters of the $\rho$, $\omega$, $\phi$, $b_1$ and $a_2$ mesons are
given by the Particle Data Group \cite{PART}.
The masses and decay widths of the $f_0$ and $f_2$ mesons are chosen to get the best
description of the experimental data \cite{MARS} for $\gamma\gamma\to\pi^0\pi^0$ process 
in the energy region of these resonances.
The parameters of the $\sigma$ meson are not known.

In the present work
the parameters of the $\sigma$ meson are found from a fit to
the experimental data \cite{MARS} for the $\gamma\gamma\to\pi^0\pi^0 $ process. 
As the $\gamma\pi$ elastic
scattering and the process $\gamma\gamma\to\pi^0\pi^0$ should be described by the common
analytical function, we use the same DRs for description of both of these processes.
In the case of reaction $\gamma\gamma\to\pi^0\pi^0$ the DRs
are saturated by the $\rho$, $\omega$ and $\phi$ mesons in the $s$ and $u$ 
channels and $\sigma$, $f_0$ and $f_2$ mesons in the $t$ channel. 
The fitting parameters are
the mass, the full width and the width of the decay into $\gamma\gamma$ of
the $\sigma$ meson. An additional parameter is the polarizabilities 
difference of $\pi^0$ meson $(\alpha-\beta)_{\pi^0}$. We consider two variants of the 
fitting: 
a) $(\alpha-\beta)_{\pi^0}$ is determined during fit procedure through the  
dispertion sum rule (DSR) \cite{RAD}; 
b) $(\alpha-\beta)_{\pi^0}$ is fixed from calculation in the framework of chiral 
perturbation theory ($\chi PT$) equal to -1.9 
(in units of $10^{-4}$fm$^3$). 
The polarizabilities sum $(\alpha+\beta)_{\pi^0}$ is calculated with help of the DSR :
$(\alpha+\beta)_{\pi^0}=0.8$.

The parameters of $\sigma$ meson obtained by the fitting in the energy region 
from $\sqrt{t}=270$ MeV up to 825 MeV ($t$ is the square of the total energy 
in $\gamma\gamma$ c.m.s.) are listed in table 1.
\begin{table}
\centering
\begin{tabular}{|c|c|c|c|c|c|} \hline
  &$m_{\sigma}$(MeV)&$\Gamma_{\sigma}$(MeV)&$\Gamma_{\sigma\to\gamma\gamma}$(keV)
&$\chi^2$&$(\alpha-\beta)_{\pi^0}$ \\ 
\hline
a &$524\pm52$&$1075\pm390$&$0.52\pm0.20$& 0.40 & -2.5 \\ \hline
b &$544\pm33$&$1170\pm285$&$0.61\pm0.13$& 0.42 & -1.9 \\ \hline
\end{tabular}
\caption{The $\sigma$ meson parameters determined by the fit}
\end{table}
As it is evident from the table, the parameters of the $\sigma$, found for the "a"
and "b" variants are in good agreement.
Fig.1 demonstrates a description of the experimental data \cite{MARS} of
the process $\gamma\gamma\to\pi^0\pi^0$ in the energy region up to 
$\sqrt{t}\approx 2$ GeV using the found values of the $\sigma$ meson parameters. 

If one takes the parameters of $\sigma$ meson obtained for the variant "b"
to evaluate the $(\alpha-\beta)_{\pi^0}$ with help of DSR \cite{RAD}, we find
$(\alpha-\beta)_{\pi^0}=-2.3$. It differs from the value (-1.9) used for fitting in 
this case. This difference could be a measure of accuracy of calculation 
of $(\alpha-\beta)_{\pi^0}$ in the framework of this DSR.

An application of the found parameters of the $\sigma$ meson for evaluation of 
the difference of $\pi^{\pm}$ meson polarizabilities with help of the DSR 
\cite{RAD} gives: 
$(\alpha-\beta)_{\pi^{\pm}}$=7.8 for the variant "a" and 8.1 for "b".
These values are bigger than ones predicted in $\chi PT$ \cite{KIR} ($\sim 5.6$),
smaller than values obtained in the frame of DSRs early \cite{RAD,PETR} (10.9--10.6)
but close to the prediction of the quark confinement
model \cite{IM} (7.05).

Fig.2 shows the results of calculation of $\gamma\pi^{\pm}$ back scattering cross 
section with help of the DRs for the fitting variants "a" (solid line) and "b" 
(doted line). The dashed line figure corresponds to $(\alpha-\beta)_{\pi^{\pm}}=0$. 
It is evident from this figure, that the value of the calculated cross 
section of $\gamma\pi^{\pm}$ scattering is practically independent of the variant
of fitting. 

For the $\gamma\pi^{\pm}$ forward scattering 
the cross section is determined by the contribution of
the Born term, the sum $(\alpha+\beta)_{\pi^{\pm}}$ and the $\rho$, 
$b_1$ and $a_2$ mesons.
The results of calculation of the forward scattering cross section 
are given in Fig.3
where the solid curver corresponds to account of
all contributions, the dashed curver shows the same contributions 
but with $(\alpha+\beta)_{\pi^{\pm}}=0$ and the dotted one is contribution 
of the Born+$(\alpha+\beta)_{\pi^{\pm}}$. 
This figure demonstrates big contribution of the sum
$(\alpha+\beta)_{\pi^{\pm}}$ in the energy region near 1GeV. 
 
The relative contributions of $(\alpha-\beta)_{\pi^{\pm}}$ and 
$(\alpha+\beta)_{\pi^{\pm}}$ into the back and
forward cross sections, respectively,
as function of the energy $\sqrt{s}$ are shown in Fig.4.
As follows from this figure the relative contributions of 
the $(\alpha-\beta)_{\pi^{\pm}}$ and $(\alpha+\beta)_{\pi^{\pm}}$ grow with 
energy and exceed 100\% in the region 
of 1GeV. This result permits to determine the pion polarizabilities
with high accuracy from the experimental data for the elastic
$\gamma\pi^{\pm}$ scattering in this energy region.

So, using the DRs we found the mass, full width and decay width
into $\gamma\gamma$ of the $\sigma$ meson from the fit to the experimental
data \cite{MARS} for the $\gamma\gamma\to\pi^0\pi^0$ process.
The analysis of the calculation of cross section of elastic 
$\gamma\pi^{\pm}$ scattering showed: 1) when extracting the $(\alpha-\beta)$ 
and $(\alpha+\beta)$
from the pion Compton scattering data it is necessary to take into
account corrections to the low energy expression;
2) the data for the elastic $\gamma\pi^{\pm}$scattering in the energy region 
up to $\sim 1$GeV (together with the $\gamma\gamma\to\pi^0\pi^0$ data) could 
be used both for determination of the pion polarizability values with high 
enough precision and for research of the $\sigma$ meson.

The authors would like to thank J. Ahrens, R. Beck, D.Drechel, A. L'vov and
Th. Walcher for helpful discussions.

\normalsize
\pagestyle{plain}
\setcounter{figure}{0}
\begin{minipage}{5.5cm}
\psfig{figure=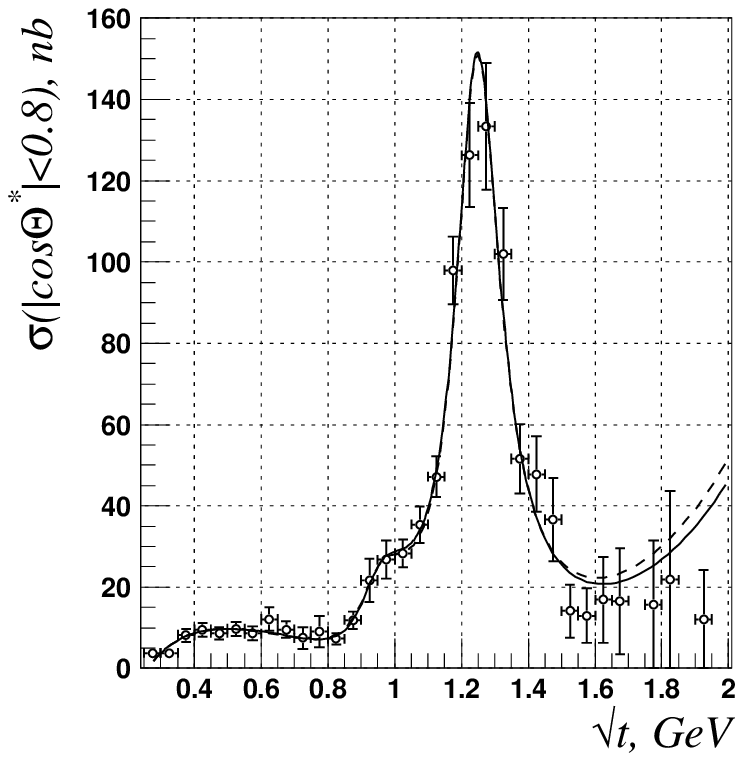,width=5.5cm}
Figure 1: Cross section for \mbox{$\gamma\gamma\to\pi^0\pi^0$}
process in energy region up to 2 GeV
\end{minipage}
\hspace*{1cm}
\begin{minipage}{5.5cm}
\psfig{figure=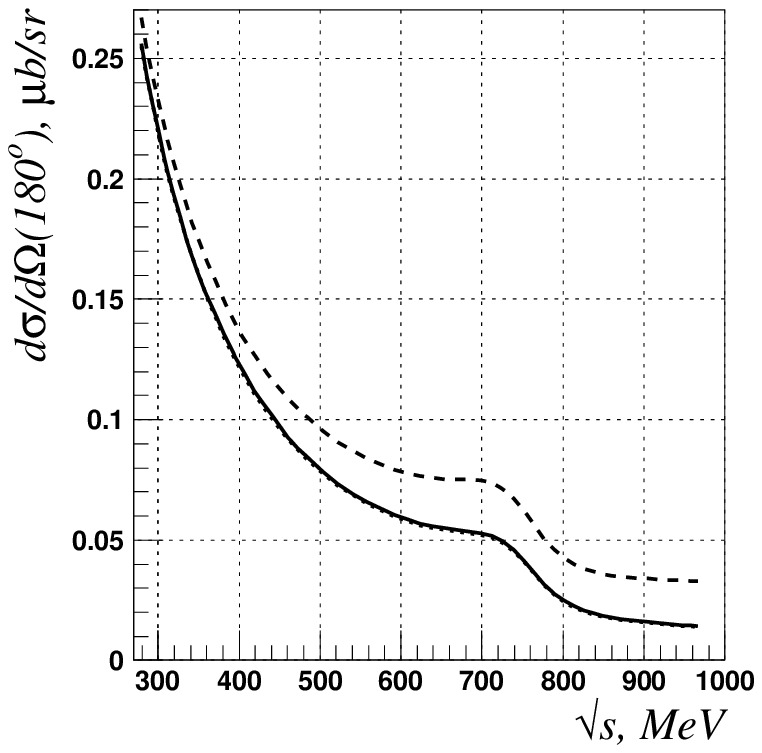,width=5.5cm}
Figure 2: Back scattering cross section
for $\gamma\pi^{\pm}\to\gamma\pi^{\pm}$ process
\\
\end{minipage}

\vspace*{1cm}
\begin{minipage}{5.5cm}
\psfig{figure=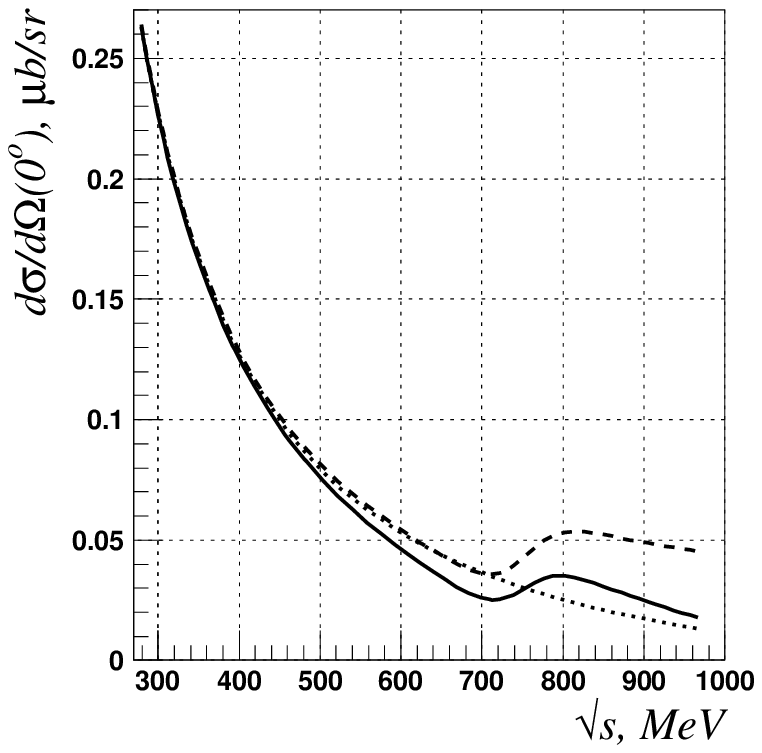,width=5.5cm}
Figure 3: Forward scattering cross section for
$\gamma\pi^{\pm}\to\gamma\pi^{\pm}$ process
\\
\end{minipage}
\hspace*{1cm}
\begin{minipage}{5.5cm}
\psfig{figure=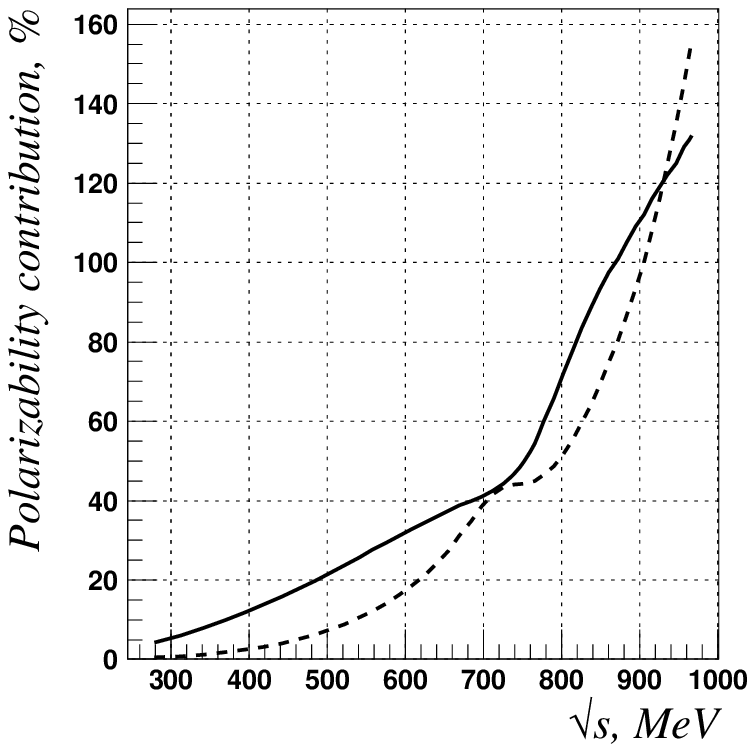,width=5.5cm}
Figure 4: The relative contributions of
$(\alpha-\beta)_{\pi^{\pm}}$
(solid line) and \mbox{$(\alpha+\beta)_{\pi^{\pm}}$} (dashed line)
\end{minipage}

\end{document}